\documentclass[11pt]{article} 
\usepackage{multirow}
\usepackage{natbib}
\usepackage[english]{babel}
\usepackage{amssymb,amsmath,amsthm}
\usepackage{color}
\usepackage[dvips]{epsfig}
\usepackage{graphics}

\usepackage[utf8]{inputenc} 

\usepackage{geometry} 
\usepackage{graphicx} 

\usepackage{booktabs} 
\usepackage{array} 
\usepackage{paralist} 
\usepackage{verbatim} 
\usepackage{subfig} 

\usepackage{fancyhdr} 
\usepackage{sectsty}
\allsectionsfont{\sffamily\mdseries\upshape} 

\usepackage[nottoc,notlof,notlot]{tocbibind} 
\usepackage[titles,subfigure]{tocloft} 

\title{Modelling slowly changing dynamic gene-regulatory networks}
\author{Antonino Abbruzzo, Ernst Wit}
\begin{document}

\maketitle

\begin{abstract}
Dynamic gene-regulatory networks are complex since the number of potential components involved in the system is very large. Estimating dynamic networks is an important task because they compromise valuable information about interactions among genes. Graphical models are a powerful class of models to estimate conditional independence among random variables, e.g. interactions in dynamic systems. Indeed, these interactions tend to vary over time. However, the literature has been focused on static networks, which can only reveal overall structures. Time-course experiments are performed in order to tease out significant changes in networks. It is typically reasonable to assume that changes in genomic networks are few because systems in biology tend to be stable. We introduce a new model for estimating slowly changes in dynamic gene-regulatory networks which is suitable for a high-dimensional dataset, e.g. time-course genomic data. Our method is based on i) the penalized likelihood with $\ell_1$-norm, ii) the penalized differences between conditional independence elements across time points and iii) the heuristic search strategy to find optimal smoothing parameters. We implement a set of linear constraints necessary to estimate sparse graphs and penalized changing in dynamic networks. These constraints are not in the linear form. For this reason, we introduce slack variables to re-write our problem into a standard convex optimization problem subject to equality linear constraints. We show that GL$_\Delta$ performs well in a simulation study. Finally, we apply the proposed model to a time-course genetic dataset T-cell.
\end{abstract}

\section{Introduction}

A single microarrays is a snapshot of the expression of genes in different samples. However, gene expression is a temporal process in which different proteins are required and synthesized for different functions and under different conditions. Even under stable conditions, due to the degradation of proteins, mRNA is transcribed continuously and new proteins are generated. This process is highly regulated. In many cases, the expression program starts by activating a few transcription factors, which in turn activate many other genes that act in response to the new condition. Transcription factors are proteins that bind to specific DNA sequences, thereby controlling the flow (or transcription) of genetic information from DNA to mRNA. For example, when cells are faced with a new condition, such as starvation \citep{natarajan2001transcriptional}, infection \citep{nau2002human} or stress \citep{gasch2000genomic}, they react by activating a new expression program. Taking a snapshot of the expression profile following a new condition can reveal some of the genes that are specifically expressed under the new condition. In order to decide the complete set of genes that are expressed under these conditions, and to discover the interaction between these genes, it is necessary to measure a time-course of expression experiments. This temporal measures allow us to determine not only the stable state following a new condition, but also the pathway and networks that were activated in order to arrive at this new state.

The biological and computational issues that are addressed when analysing gene expression data in general, and time-course expression data in particular, can be presented using four analysis levels: experimental design, data analysis, pattern recognition and networks. A review paper on each of these four analysis levels is given by \cite{bar2004analyzing}.

In this paper we focus on estimating sparse slowly dynamic network estimation with Graphical models.
Graphical models explore conditional independence relationships between random variables. We can divide graphical models into directed graphical models, e.g. Bayesian networks  \citep{jensen1996introduction, neapolitan2004learning} and undirected graphical models, e.g. Gaussian graphical models \citep{whittaker1990graphical, lauritzen1996graphical}.

Bayesian networks (BNs) have been used to estimate the structure between multiple interaction quantities such as expression levels of different genes \citep{friedman2000using}. However, Bayesian networks suffer two major limitations. Firstly, no cycle can be estimated, secondly BNs perform poorly on sparse microarray data as shown by \cite{husmeier2003sensitivity}. The first limitation of BN can be overcome. Dynamic Bayesian networks (DBNs) have been proposed to estimate directed graphs with cycles \citep{murphy2002dynamic, ghahramani1998learning, perrin2003gene}. In other words, DBNs are an extension of Bayesian networks which have the advantage that cycle can be inferred. However, they can only estimate directed links. Instead, we need to estimate both directed and undirected links for time-course genetic dataset. Gaussian graphical models can be adapted to estimate graphs with directed and undirected links. The main advantage for GGMs is that the precision matrix, i.e. the inverse of the covariance matrix, represents the conditional independence.

The literature on estimating an inverse covariance matrix goes back to \cite{dempster1972covariance}, who advocated the estimation of a sparse dependence structure, i.e., setting some elements of the inverse covariance matrix to zero. The complexity of the covariance matrix is reduced when elements in the inverse of this matrix are fixed at zero. Moreover, it has been shown that most of the networks in biology are sparse, which means that most of the elements in the precision matrix are equal to zero. The standard approach in statistical modelling to identify zeros in the precision matrix is  the backward stepwise selection method, which starts by removing the least significant edges from a fully connected graph, and continues removing edges until all remaining edges are
significant according to an individual partial correlation test.
This procedure does not work in the case of multiple testing, i.e. many of the links will be estimated to be different from zero when they are not and vice versa. A conservative simultaneous testing procedure was proposed by \cite{drton2004model}. However, \cite{breiman1996heuristics} showed that this two-step procedure, in which parameter estimation and model selection are done separately, can lead to
instability. Such instability means that small changes in the dataset or small perturbations result in completely different estimated graph structures \citep{breiman1996heuristics}.

The idea of \cite{tibshirani1996regression}, which has been extensively and successfully applied in regression models, can be used to estimate sparse graphs, i.e. to induce zeros in the estimated inverse covariance matrix. This idea is based on the $\ell_1$-norm penalty, i.e. the sum of the absolute values of the inverse of the covariance matrix has to be less than or equal to a tuning parameter. The smaller the tuning parameter is, the more zero will be estimated in the precision matrix. \cite{banerjee2008model}, \cite{meinshausen2006high}, \cite{d2006first} proposed penalized graphical models with $\ell_1$-norm penalty. However, penalized graphical models estimate a single Gaussian graphical model while in many applications it is more realistic to fit a collection of such models, due to the different experimental conditions across time points. The graphical lasso \citep{friedman2008sparse} can be used to estimate a network for each time point. However, the application of this graphical model results in $t$ static networks estimation, where $t$ is the number of time points. In order to to discover a common structure and jointly estimate common links across graphs, \cite{guo2011joint} proposed a method that links the estimation of separate graphical models through a hierarchical penalty. This graphical model leads to improvements compared to fitting separate models, since it borrows information from other related graphs.
Recently, \cite{wit2012} proposed factorially coloured graph to estimate this common structure. The idea is to combine sparsity and colouring to built several possible models. The latter two models are still not able to estimate the dynamic changes of the network.

In this paper we propose a model to estimate dynamic graphs with slow changing using $\ell_1$-regularization framework. The main idea is to impose $\ell_1$-norm to penalize changing in the networks over time. Moreover, an $\ell_1$-norm penalty is imposed on the precision matrix to induce sparsity in the graph. The new model, called GL$_\Delta$, is suitable for studying high-dimensional dataset. In other words, we consider a penalized likelihood estimation problem subject to linear constraints, which are necessary to induce slow changing in the dynamic graphs. In order to solve this penalized maximum likelihood problem, we need to fix two tuning parameters that regulate sparsity and penalize changing in the dynamic network. For this reason, we propose a heuristic search method to find two optimal values for the tuning parameters. We take advantage of an efficient solver developed by \cite{wang2009solving} to solve the optimization problem with linear constraints.

The rest of this paper is organized as follows. In the next section, we give a description of our motivating example and a brief overview of Gaussian graphical models. In Section \ref{method}, we describe the slowly changing dynamic network. In Section \ref{results}, we show the results of a simulation study and apply GL$_\Delta$ to the time-course genetic dataset T-cell. Finally, we discuss the advantages of our method and point out further directions for development.

\section{Motivation: T-cell activation}
An important issue in system biology is to understand the system of interactions among several biological components such as protein-protein interaction and gene regulatory networks. Hence, several techniques have been developed to collect data from different organisms. For instance, microarrays measure gene expression levels, i.e. the concentration of the messenger RNA produced when the gene is transcribed. 	

A time-course genetic dataset \lq\lq{}T-cell\rq\rq{} is our motivating example. The aim of the experiment was to collect temporal data to identify the underlying gene regulatory networks.

Two cDNA microarray experiments were performed to collect gene expression levels for T-cell activation analysis. Activation of T-Cell was produced by stimulating the cells with two treatments: the calcium ionosphere and the PKC activator phrorbolester PMA. The human T-cells coming from the cellular line Jakart were cultured in a laboratory. When the culture reached a consistency of $10^6$ cells/ml, the cells were treated with the two treatments PMA and PKC. Gene expression levels for 88 genes were collected for the following times after the treatments: 0, 2, 4, 6, 8, 18, 24, 32, 48, 72 hours. In the first experiment the microarray was dived such that 34 sub-array were obtained. Each of these 34 sub-arrays contained the strands of the 88 genes under investigation. Strands are the complementary base for the rRNA, which is the transcribed copy of a single strand of DNA after the process of transcription. In the second microarray experiment the microarray was dived into 10 sub-arrays.  Each of these 10 sub-arrays contained the strands of the 88 genes under investigation. Each microarray was composed by ten different slides which were used for the two experiments to collect temporal measurements.  For example for time 0 a set of cells were hybridized in the first slide before the treatments were considered and after the cell cultured reached the right density. For the second time point (time 2), another set of cells were hybridized in a second slide. The experiment was conducted by \citep{rangel2004modeling}.

At this point we assume that the technical replicates are independent samples, and that the temporal replicates are dependent replicates from the same samples. These two assumptions result in a dataset with 44 independent replicates across 10 time points. These are strong assumptions. This means that the conclusion on the analysis on T-cell should be critically considered.

Two further steps were conducted by \cite{rangel2004modeling} to obtain a set of genes that were highly expressed and normalized across the two microarrays. Firstly, genes with high variability between the two microarryes and within the same time point were removed. No further information is present in the paper about for example the minimum level of reproducibility they adopted. According to \cite{rangel2004modeling}, thirty genes have to be removed since they do not showed enough reproducibility. Secondly, a normalization methods were applied to remove systematic variation due to experimental artifacts. The normalization method used by \cite{rangel2004modeling} is described in the paper written by \cite{bolstad2003comparison}.

\section{Methods}
\label{method}
In this section, we describe the tools that we need in order to study the underlying time-varying genomic network for the T-cell data. We believe that time-course datasets should be analysed exactly in this way, whereas there was no point in collecting time-course data. The tools should be adjusted to the needs of the biologist that wants to infer particular aspects of the system. Firstly, we introduce a graphical model. Secondly, we extend this model to slowly changing graphical lasso. 

\subsection{Gaussian graphical model}
A graphical model $(G,\mathbb{P})$, where $\mathbb{P}$ is a multivariate normal distribution $N(\boldsymbol \mu, \boldsymbol \Sigma)$ with mean $\boldsymbol \mu$ and variance covariance matrix $\boldsymbol \Sigma$, is called a Gaussian graphical model or a covariance selection model \citep{dempster1972covariance}. Let $\boldsymbol \Theta = \boldsymbol \Sigma^{-1}$ be the precision or concentration matrix, then $\boldsymbol \Theta$ contains all conditional independence information for the Gaussian graphical model. In fact, if $\theta_{ij} = 0$ then $Y_i$ is independent of $Y_j$ given the rest, i.e. the pairwise Markov property $Y_i \perp Y_j|\mathbf{Y}_{V\setminus\{i,j\}}$. In fact, it can be shown that given the set of $\theta_{ij} = 0$, a joint normal probability distribution $f(\mathbf{y})$  can be factorized as a product of functions $f$ which do not jointly depend to $y_i$ and $y_j$ when $\theta_{ij} = 0$.
Gaussian graphical models fail when the number of observation is fewer than the number of variables. This situation is really common for dynamic biological networks. Moreover, dynamic biological networks are sparse, which means that the number of links is small with respect to the possible number of connections. Not only sparsity is our current bet knowledge of genomic interaction but it is also computational useful.

\subsection{Graphical lasso for slowly changing dynamic networks}
\label{Graphical lasso for slowly changing dynamic networks estimations}

The main aim of this paper is to show that we can use the idea of penalized likelihood to estimate sparse dynamic networks with slow changes across time points. A graph with few edges is sparse, and a graph with many edges is dense. Formally, a graph $G = (V, E)$ is said to be sparse if $ \bar{\bar{E}} = O(\bar{\bar{V|}})$, where $\bar{\bar{V}}$ is the number of vertices and $\bar{\bar{E}}$ is the number of links (couple of vertex or node). A graph $G$ is said to be dense if $\bar{\bar{E}} = O(\bar{\bar{V}}^2)$. These two definitions are given by \cite{preiss2008data}. Roughly speaking, high dimensional statistical inference is possible, in the sense of leading to reasonable accuracy or asymptotic consistency, if $\bar{\bar{E}} \mbox{log}(\bar{\bar{V}}) << n$, where $n$ is the number of observations. In other words, accuracy and consistency of the results depend on how one define sparsity.

Much of the methodology in high-dimensional analysis relies on the idea of penalizing the $\ell_1$-norm of the precision matrix $\boldsymbol \Theta$, i.e. $\sum_{i=1}^{p} \sum_{j=1}^p |\theta_{i,j}| \leq \rho$, for $i > j$ where $\rho$ is a tuning parameter that regulates the sparsity. The smaller the value of the tuning parameter $\rho$ is the most sparse is the estimated matrix $\boldsymbol \Theta$. Such $\ell_1$-penalization has become tremendously popular due to its computational attractiveness (i.e. convex function) and its statistical properties which are optimal under certain conditions; mainly, if we want to minimize a prediction error while we are choosing a model as simple as possible. It is important that model selection and parameter estimation are done contemporaneously. In fact, \cite{breiman1996heuristics} showed that using hypothesis testing for model selection this two steps procedure brings instability in the model. Instability means that if we slightly perturb the data, then the results can change considerably. Whereas, $\ell_1$-penalized methodologies allow us to do model selection and estimation, simultaneously.
Moreover, our main idea is to impose the $\ell_1$-norm to penalize changes in the networks through time points.

Given a dynamic graph $G = (V, E)$, which is graph where the same nodes are measured across a finite number of time points, the edge set $E$ can be partitioned into natural partitions $S_s, N_s$ which are shown in the matrix $\boldsymbol \Theta$. The natural partitions $S_s$ and $N_s$ are interpretable as self-self interactions at lag $s$ and networks interactions at lag $s$. Each of this subset can be further partitioned and we indicate with $S_s^t$ and $N_{s}^{t}$ these new sub-partitions. $S_s^t$ is the self-self term at lag $s$ and time $t$ and $N_{s}^{t}$  is the network at lag $s$ and time $t$. Let us consider a Gaussian graphical model $M = (G, \mathbb{P})$, where $\mathbb{P}$ is a multivariate normal distribution parametrized by $\boldsymbol \Sigma^{-1} = \boldsymbol \Theta$, then we consider the following decomposition of the precision matrix $\boldsymbol \Theta$: 
\[\boldsymbol \Theta = \left[ \begin{array}{cc|cc|cc|cc}
            S_{0}^{1} & N_{0}^{1} & S_{1}^1 & N_{1}^1 & S_{2}^1 & N_{2}^1 & \ldots & \ldots \\
                     & S_{0}^1& N_{1}^1 & S_{1}^1 & N_{2}^1 & S_{2}^1 & \ldots & \ldots \\
            \hline
                     &         & S_{0}^2 & N_{0}^2 & S_{1}^2 & N_{1}^2 & S_{2}^2 & N_{2}^2\\
                     &         &          & S_{0}^2& N_{1}^2 & S_{1}^2 & N_{2}^2 & S_{2}^2\\
            \hline
                     &         &          &         & S_{0}^3 & N_{0}^3 & S_{1}^3 & N_{1}^3 \\
                     &         &          &         &          & S_{0}^3 & N_{1}^3 & S_{1}^3 \\
            \hline
                     &         &          &         &          &         & \ddots   & \vdots\\
                     &         &          &         &          &         &          & \ddots \end{array}
            \right], \]
where $S_{s}^t$ are self-self conditional correlations of the genes across time lag $s$ and time $t$, and $N_{s}^t$ is a genetic network with time lag $s$ and time $t$. Our interest is in detecting evolution of the networks, where the evolution is evaluated from the element-wise differences between $N_{s}^ t$ and $N_{s}^{t+1}$, i.e
\[
|N_{s}^t| - |N_{s}^{t+1}|,
\]
where $|\cdot|$ indicates the absolute value. 

Our aim is to estimate \lq\lq significant\rq\rq{} differences between these elements while the general structure is still sparse.

\subsection{Maximum likelihood estimation}
\label{sec:Maximum likelihood estimation for delta graphical lasso.}
Suppose that $\mathbf{Y}^{(1)}, \ldots, \mathbf{Y}^{(n)}$ with $\mathbf{Y}^{(i)} \in \mathbb{R}^{gt}$, where $g$ is the number of random variables per each time point and $t$ is the number of time points, are independent and identically distributed  as a multivariate normal distribution with mean $\mathbf{0}$ and variance $\boldsymbol \Sigma$. This optimization problem can be written in the following standard form:
\begin{eqnarray}
\label{eq:copulaloglikpen3431}
\nonumber
\hat{\boldsymbol \Theta} &:=& \mathop{\mbox{argmin}}_{\boldsymbol \Theta}\{-\log (\mbox{det}(|\boldsymbol \Theta|)) +\mbox{tr}(\mathbf{S} \boldsymbol \Theta) +\\
 & &  \lambda_1 \mathbf{x}^+ + \lambda_1 \mathbf{x}^- \} \\
\nonumber
\mbox{subject to}  &&\mathbf{B}(\boldsymbol \Theta) - \mathbf{x}^+ + \mathbf{x}^- =  \mathbf{0}\\
\nonumber
&&\boldsymbol \Theta \succ 0, \mathbf{x}^+, \mathbf{x}^- \geq \mathbf{0} .
\end{eqnarray}
where $\mbox{det}(\cdot)$ is the determinant, $\mbox{tr}(\cdot)$ indicates the trace, $\mathbf{B}(\boldsymbol \Theta)$ indicates the usual $\ell_1$ constraint, i.e. $||\boldsymbol \Theta||_1 \leq \rho_1$. Note that $\mathbf{x}^+$ and $\mathbf{x}^-$ are slack variables in $\mathbb{R}^m$ where $m = gt(gt-1)/2$. Here $\lambda_1$ is a smoothing parameter, which regulates the sparsity in the precision matrix $\boldsymbol \Theta$ for sparse Gaussian graphical models.

Now we want to penalize the difference between networks with lag $s$ at time $k$ and the same networks at time $k+1$, i.e.
\begin{eqnarray}
\label{eq:deltacontr}
||\Delta \boldsymbol \Theta||_1 &=& \sum_{s = 0}^{t-1} \sum_{k= 0}^{t-1} ||N_{s}^{k}-N_{s}^{k+1}||_1 =\\
\nonumber
& & \sum_{s=0}^{t-1} \sum_{k=0}^{t-1} \sum_{i,j}|\theta_{(i,k),(j,k+s)} - \theta_{(i,k+1),(j,k+1)} | \leq \rho2.
\end{eqnarray}
In order to take advantage of LogDetPPA, we need to built a linear map $\mathbf{A}$ such that a system of linear equations is included in the optimization problem (\ref{eq:copulaloglikpen3431}). We aim to write 
$\mathbf{A}(\mathbf{\Theta}) $ such that $\mathbf{A}(\mathbf{\Theta}) \equiv ||\Delta \boldsymbol \Theta||_1$. 
LogDetPPA can only manage to equality constraint but (\ref{eq:deltacontr}) represents inequality constraints so it needs to be converted. For this reason, we introduce another vector of slack variables $\mathbf{y}^+, \mathbf{y}^-$ such that:
\begin{eqnarray}
\label{eq:deltacontr2}
\nonumber
\sum_{s=0}^{t-1} \sum_{k=0}^{t-1} \sum_{i,j}|\theta_{(i,k),(j,k+s)} -
\theta_{(i,k+1),(j,k+1)} | - y^+_{k} + y^-_{k} = 0
\end{eqnarray}
where $k = 1, \ldots, K$, and $\mathbf{y}^+, \mathbf{y}^- \geq 0$. The optimization problem (\ref{eq:copulaloglikpen3431}) is now written as:
\begin{eqnarray}
\label{eq:copulaloglikpen34311}
\nonumber
\hat{\boldsymbol \Theta} &:=& \mathop{\mbox{argmin}}_{\boldsymbol \Theta}\{-\log(\mbox{det}(\boldsymbol \Theta))+ \mbox{tr}(\mathbf{S} \boldsymbol \Theta) +\\
 &&\lambda_1 \mathbf{x}^+ + \lambda_1 \mathbf{x}^- + \lambda_2 \mathbf{y}^+ + \lambda_2 \mathbf{y}^- \}\\
\nonumber
&&\\
\mbox{subject to}  &&\mathbf{B}(\boldsymbol \Theta) - \mathbf{x}^+ + \mathbf{x}^- =  \mathbf{0}\\
\label{eq:123}
\nonumber
  &&\mathbf{A}(\boldsymbol \Theta) - \mathbf{y}^+ + \mathbf{y}^- =  \mathbf{0}\\
\nonumber
&&\boldsymbol \Theta \succ 0, \mathbf{x}^+, \mathbf{x}^-,  \mathbf{y}^+, \mathbf{y}^- \geq \mathbf{0}.
\end{eqnarray}
The optimization problem (\ref{eq:deltacontr2}) subject to (4) is a convex optimization problem with linear constrains which can be solved by using LogDetPPA.

It should be notice that both $\lambda_1$ and $\lambda_2$ are non-negative smoothing parameters that need to be selected.  We consider a grid of values ($\lambda_1, \lambda_2$) and minimize information criterion scores such as AIC, AICc, and BIC. Then we use stability selection to select a more stable graph \citep{meinshausen2010stability}.

\paragraph{Example: T-cell} We apply GL$_{\Delta}$ to T-cell dataset, where 4 genes and 2 time points were considered, in order to show a small example. Table \ref{concentrationooo} shows the estimated precision matrix. Here, we fixed the tuning parameters $\lambda_1 = 0.01$ and $\lambda_2 = 0.1$.

\begin{table}[h!]
\centering
\begin{tabular}{rrrrrr|rrrr}
 Time &\multicolumn{5}{c}{ \em 1} &\multicolumn{4}{c}{ \em 	2}  \\
 &  Gene & \em ZNF &\em CCN &\em SIV &\em SCY & \em ZNF &\em CCN &\em SIV &\em SCY \\
  \hline
\multirow{4}{*}{\em 1} & \em ZNF & 1.24 & 0.00 & -0.26 & 0.18 & -0.22 & -0.11 & -0.11 & -0.07 \\
&\em CCN & - & 1.49 & 0.00 & -0.17 & -0.18 & -0.84 & 0.06 & 0.12 \\
&\em SIV & - & - &  1.44 & 0.00 & -0.15 & 0.08 & -0.69 & -0.01 \\
&\em SCY & - & - & - & 1.19 & 0.02 & 0.13 & 0.41 & -0.10 \\
\hline
\multirow{4}{*}{\em 2} & \em ZNF  & - & - & - & - & 1.07 & -0.02 & 0.00& 0.12 \\
&\em CCN  & - & - & - & - & - & 1.55 & 0.00& 0.24 \\
 &\em SIV & - & -& - & - & - & -& 1.52 & 0.00 \\
&\em SCY & - & - & - & - & - & - &- &  1.08 \\
\end{tabular}
  \caption{\label{concentrationooo}Conditional covariance $\hat{\boldsymbol \Theta}$ based on 44 replicates for 4 genes measured across 2 time points. The tuning parameters $\lambda_1$ and $\lambda_2$ were fixed to 0.01 and 0.1, respectively.}
\end{table}

Let us focus on the differences between elements of network at lag 0 at time 1 and at time 2. Then Table \ref{concentrationooo} shows that "significant" differences were estimated between ZNF-CCN and ZNF-SIV. In fact, an edge was absent between ZNF and CCN at time 1 but it was present at time 2. Opposite is the situation for ZNF-SIV.

\section{Results}
\label{results}
\subsection{Simulation study for delta graphical lasso model}
\label{sec:Simulation study for delta graphical lasso model}
We considered a simulation study to show the performance of the proposed model. Tab. \ref{tab:simulstudyscheme1} shows the simulation study scheme in which four different scenarios are considered. Here for different scenarios we mean that the number of nodes, links or time points change while the structure of the networks is the same.
\begin{table}[!h]
\centering
\fbox{%
\begin{tabular}{*{6}{c}}
 &g& t&  gt & n\\
\hline
1& 20& 3 & 60&50 \\
2 & 40& - & 120& -\\
3 & 60& - & 180& -\\
4 & 80& - & 240& -\\
\end{tabular}}
\caption{\label{tab:simulstudyscheme1}Simulation study scheme in which four scenarios are represented. The first column is an identification number, the second one indicates the number of variables per each time point (third column). The number of independent samples are represented in the last column.}
\end{table}

For each scenario we simulate 100 datasets from a multivariate normal distribution with $\boldsymbol \mu$ equal to zero and $\boldsymbol \Sigma$ equal to the inverse of a precision matrix $\boldsymbol \Theta$. The structure of the graph slowly changes across time. In fact, we want to consider graph with similar structures across some time points. Let us consider a graph with $gt \times gt$ nodes and $m$ connections and let's say that these $g$ nodes are observed at $t$ time points. In order to build our matrix $\boldsymbol \Theta$ we start to build $N_{0}^{1}$, i.e. the network at lag 0 and time point 1. We can refer to this network as the starting point network. Then for $N_{0}^{2}$ we assume that few changes happened so that $N_{0}^{1}$ and $N_{0}^{2}$ are similar networks. For example we allow $n_1$ edges to be birth and $n_2$ edges to be death. We repeat this procedure for $t-1$ times and then we put the $N_{0}^i$ sub-matrices into the matrix $\boldsymbol \Theta$. Note that we assume networks in which $y_{i,k-1}$ and $y_{j,k+1}$ are independent given $y_{.,k}$ so that network with lag greater than 1 are filled with zeroes. We increase the number of nodes in the graph from scenarios 1 to 4. Random variables associated with these added nodes are independent. We keep the number of replicates and time points constants. Note that, the number of replicates is fewer than the number of random variables.

We take advantage of the R package $\tt{simone}$ to simulate networks with few changing points. The function \verb coNetworks \ gives the opportunity to create such structures with $n = n_1+n_2$ links different from a given structure. Note that we have implemented the constraints and used $\tt{R.Matlab}$ to connect Matlab and R.
\begin{table}[!h]
\begin{center}
\begin{tabular}{|rr|rrrr|}
  \hline
   & &  $\bar{FP}$ & $\bar{FN}$ & $\bar{FD}$ & $\bar{FnD}$ \\
  \hline

 &\color{red}{AICc} & 0.0092 & 0.0811 & 0.2000 & 0.0031  \\
 1&BIC & 0.0363 & 0.0139 & 0.4873 & 0.0005  \\
 &AIC & 0.0698 & 0.0069 & 0.6470 & 0.0003  \\

  \hline
 &\color{red}{AICc}  & 0.0057 & 0.0447 & 0.2899 & 0.0006  \\
2 &BIC & 0.0088 & 0.0321 & 0.3826 & 0.0005  \\
 &AIC & 0.0437 & 0.0041 & 0.7514 & 0.0001  \\

   \hline
  &\color{red}{AICc} & 0.0016 & 0.4585 & 0.2730 & 0.0036  \\
 3& BIC  & 0.0016 & 0.4585 & 0.2730 & 0.0036  \\
  &AIC  & 0.0288 & 0.1452 & 0.8088 & 0.0012  \\

\hline
& \color{red}{AICc} & 0.0091 & 0.1034 & 0.1680 & 0.0052  \\
4&  BIC & 0.0396 & 0.0517 & 0.4527 & 0.0027  \\
&AIC  & 0.0670 & 0.0000 & 0.5704 & 0.0000 \\
\hline
\end{tabular}
\end{center}
\caption{\label{tab:simulstudyfordelta} The average of the proportions of how many links have been correctly estimated were calculated by the False Positive (FP), False Negative (FN), False Discovery (FD) and False not Discovery (FnD).}
\end{table}
Table \ref{tab:simulstudyfordelta} shows the average of false positive, false negative and false discovery after 100 simulation were run.
These results show that the model is reliable and it can be used for real applications when few changes in different time points are present. We typically prefer to use the AICc to select the model of interest.

\subsection{Application to T-cell}
\label{sec:Application to T-cell}

\begin{figure}
\includegraphics[scale=0.8]{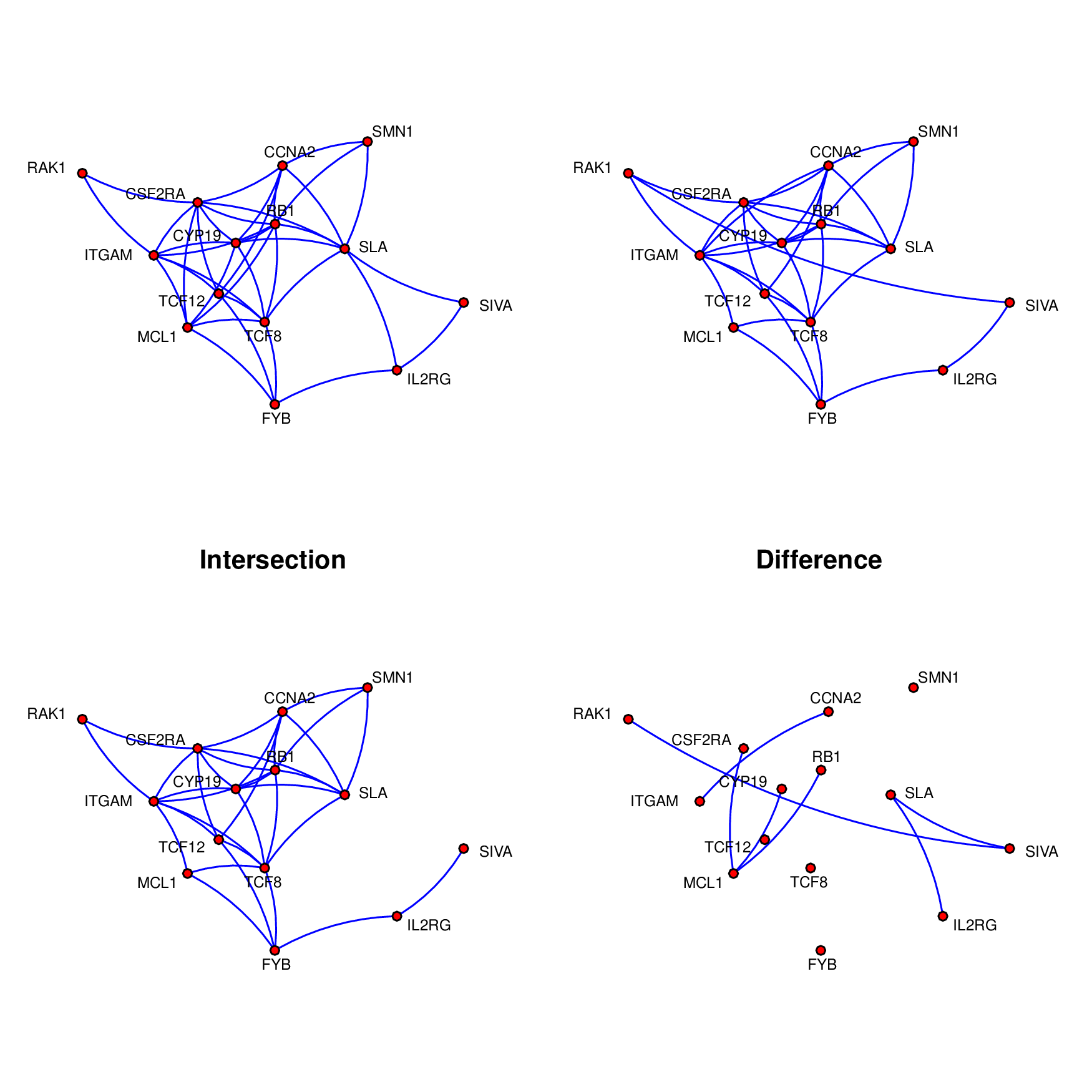}
\caption{Graph, intersection and difference between time 1 and time 2}\label{fig:diffres1}
\end{figure}

\begin{figure}
\includegraphics[scale=0.8]{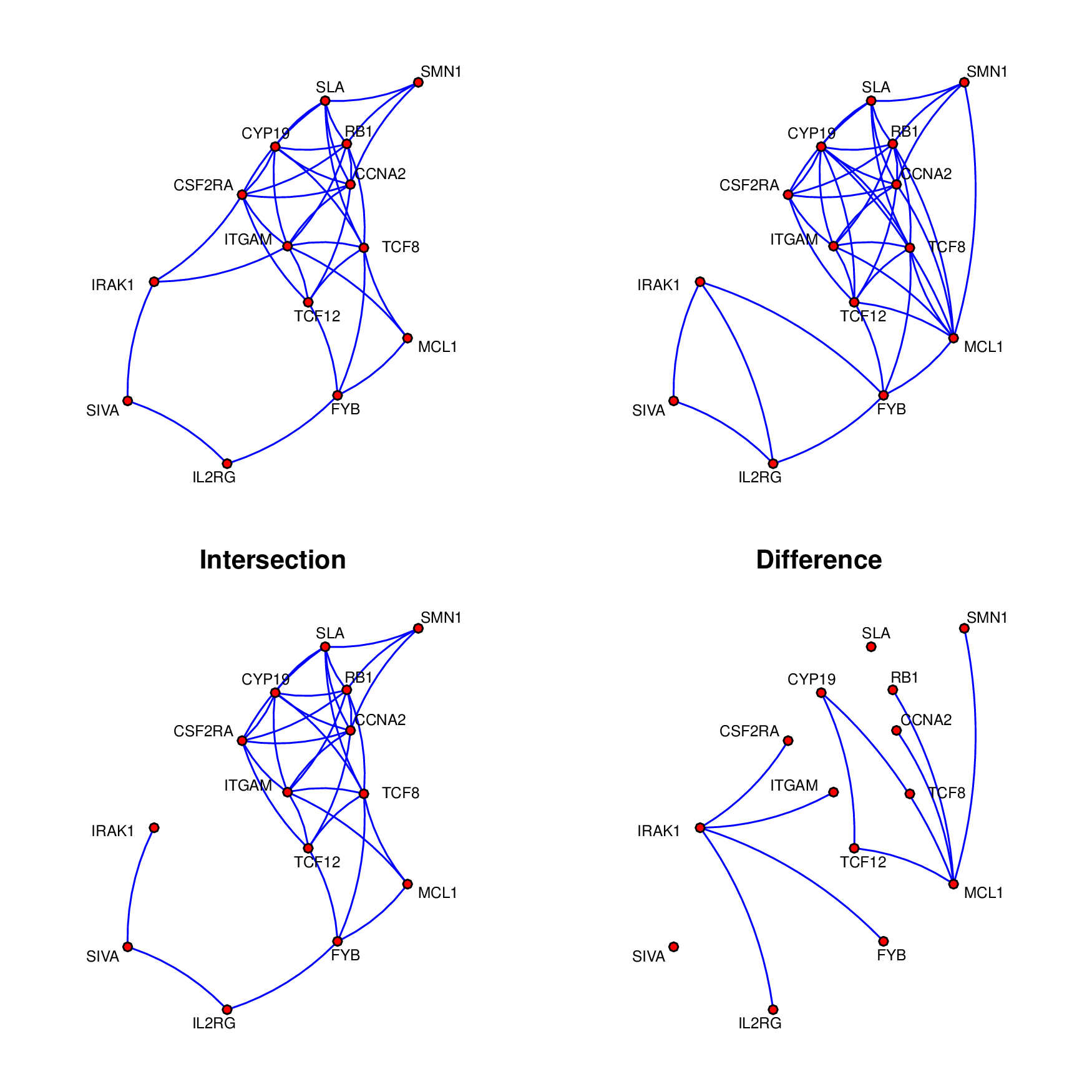}
\caption{Graph, intersection and difference between time 2 and time 3}\label{fig:diffres2}
\end{figure}

\begin{figure}	
\includegraphics[scale=0.8]{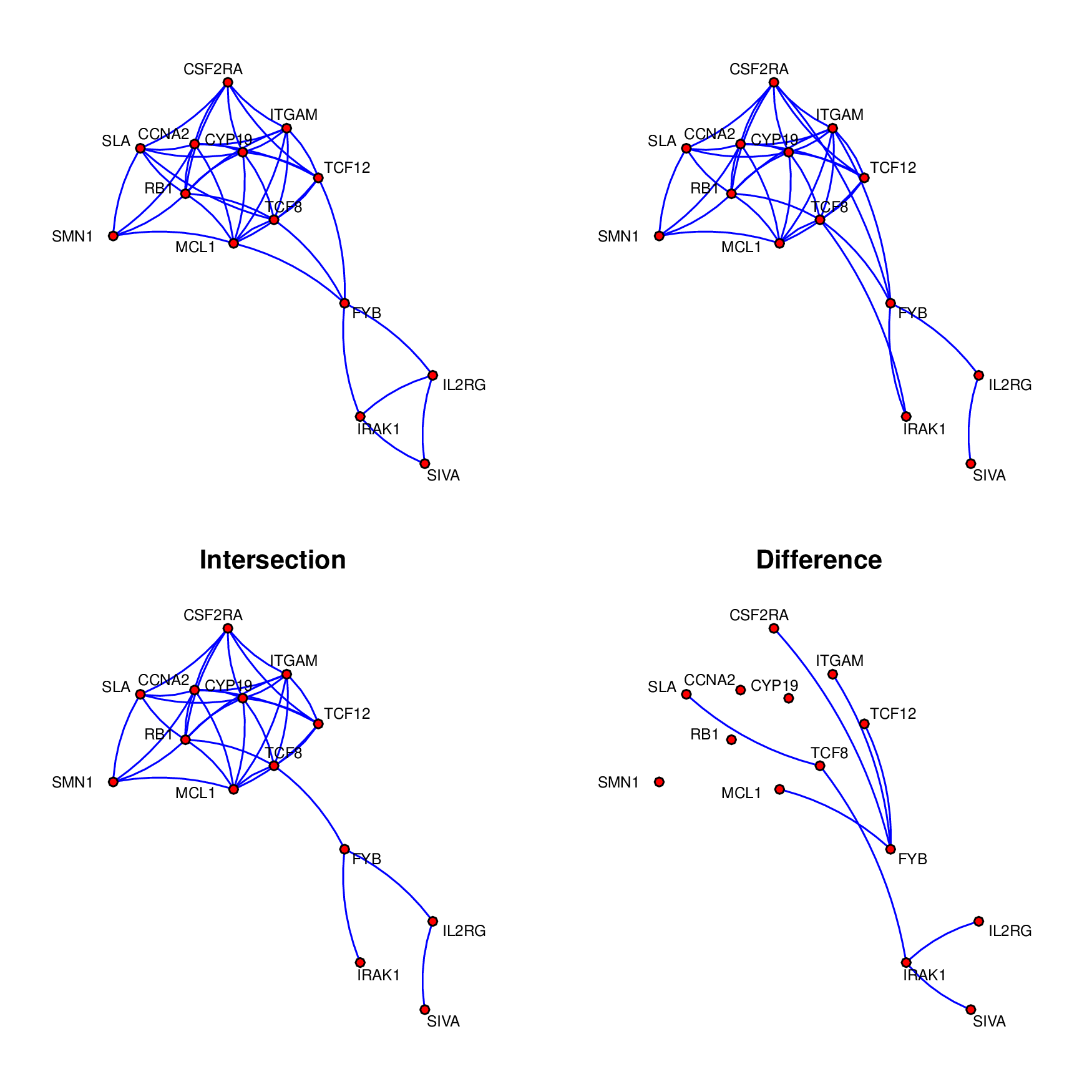}
\caption{Graph, intersection and difference between time 3 and time 4}\label{fig:diffres3}
\end{figure}

In this subsection we apply the proposed model GL$_\Delta$ to a real data set human T-cell dataset. We assume that genes which are two time points apart, i.e.  $Y_{s,t}$ and $Y_{s,t+2}$, are conditional independent given the intervening observations. This means that the edge set for networks at lag 2, i.e. $N_2$, is an empty set. Figures \ref{fig:diffres1}, \ref{fig:diffres2} and \ref{fig:diffres3} are obtained from the estimation procedure where two graphs (upper-left, upper right), intersection (bottom-left)  and difference (bottom-right) between time 1 and time 2, time 2 and time 3, and time 3 and time 4 are represented. We note that whereas initially MCL1, a pro-survival BCL2 family member, is a highly connected node in the T-cell network. It is known that SCF(FBW7) regulates cellular apoptosis by targeting MCL1 for ubiquitylation and destruction \citep{Inuzuka11}. This is probably why over the first 4 time-points it increasingly loses  connections to other genes. 

Once a graph has been estimated and changes have been evaluated, other questions on how to analyze time-evolution networks might be posed. In what way do new entities enter a network? Does the network retain certain graph properties as it grows and evolves? Does the graph undergo a phase transition, in which its behaviour suddenly changes? In answering these questions it is of interest to have a diagnostic tool for tracking graph properties and noting anomalies and graph characteristics of interest. For example, a useful tool is ADAGE \citep{mcglohon2007adage}, which is a software package that analyzes the number of edges over time, number of nodes over time, densification law, eigenvalues over increasing nodes, size of largest connected component versus nodes, number of connected components versus nodes and time, comparative sizes of connected components over time.


\section{Conclusion}
We have shown in this paper that representing genomic interactions like static graphs is particularly unsuitable for answering important biological questions about the behaviour of an genomic system over a particular time-course. Human t-cell data were used to study the developmental aspects of the sparse genomic interactions and one important result, backed up by recent research, is that MCL1 is targeted early on and thereby loses its connections to the rest of the genomic network. We use a sparse dynamic graphical model to infer these slowly changing networks. One of the major contributions is that this methodology is capable of providing fast inference about the dynamic network structure in moderately large networks. Until now, even sparse static inference could be painstakingly slow and would typically lack obvious interpretation. 

\bibliographystyle{natbib}
\bibliography{MyCollection1}

\begin{thebibliography}{}

\bibitem[Banerjee {\em et~al.}(2008)Banerjee, El~Ghaoui, and
  d'Aspremont]{banerjee2008model}
Banerjee, O., El~Ghaoui, L., and d'Aspremont, A. (2008).
\newblock Model selection through sparse maximum likelihood estimation for
  multivariate gaussian or binary data.
\newblock {\em The Journal of Machine Learning Research\/}, {\bf 9}, 485--516.

\bibitem[Bar-Joseph(2004)Bar-Joseph]{bar2004analyzing}
Bar-Joseph, Z. (2004).
\newblock Analyzing time series gene expression data.
\newblock {\em Bioinformatics\/}, {\bf 20}(16), 2493--2503.

\bibitem[Bolstad {\em et~al.}(2003)Bolstad, Irizarry, {\AA}strand, and
  Speed]{bolstad2003comparison}
Bolstad, B., Irizarry, R., {\AA}strand, M., and Speed, T. (2003).
\newblock A comparison of normalization methods for high density
  oligonucleotide array data based on variance and bias.
\newblock {\em Bioinformatics\/}, {\bf 19}(2), 185--193.

\bibitem[Breiman(1996)Breiman]{breiman1996heuristics}
Breiman, L. (1996).
\newblock Heuristics of instability and stabilization in model selection.
\newblock {\em The Annals of Statistics\/}, {\bf 24}(6), 2350--2383.

\bibitem[d'Aspremont {\em et~al.}(2006)d'Aspremont, Banerjee, and
  Ghaoui]{d2006first}
d'Aspremont, A., Banerjee, O., and Ghaoui, L. (2006).
\newblock First-order methods for sparse covariance selection.
\newblock {\em Arxiv preprint math/0609812\/}.

\bibitem[Dempster(1972)Dempster]{dempster1972covariance}
Dempster, A. (1972).
\newblock Covariance selection.
\newblock {\em Biometrics\/}, pages 157--175.

\bibitem[Drton and Perlman(2004)Drton and Perlman]{drton2004model}
Drton, M. and Perlman, M. (2004).
\newblock Model selection for gaussian concentration graphs.
\newblock {\em Biometrika\/}, {\bf 91}(3), 591--602.

\bibitem[Friedman {\em et~al.}(2008)Friedman, Hastie, and
  Tibshirani]{friedman2008sparse}
Friedman, J., Hastie, T., and Tibshirani, R. (2008).
\newblock Sparse inverse covariance estimation with the graphical lasso.
\newblock {\em Biostatistics\/}, {\bf 9}(3), 432.

\bibitem[Friedman {\em et~al.}(2000)Friedman, Linial, Nachman, and
  Pe'er]{friedman2000using}
Friedman, N., Linial, M., Nachman, I., and Pe'er, D. (2000).
\newblock Using bayesian networks to analyze expression data.
\newblock {\em Journal of computational biology\/}, {\bf 7}(3-4), 601--620.

\bibitem[Gasch {\em et~al.}(2000)Gasch, Spellman, Kao, Carmel-Harel, Eisen,
  Storz, Botstein, and Brown]{gasch2000genomic}
Gasch, A., Spellman, P., Kao, C., Carmel-Harel, O., Eisen, M., Storz, G.,
  Botstein, D., and Brown, P. (2000).
\newblock Genomic expression programs in the response of yeast cells to
  environmental changes.
\newblock {\em Molecular biology of the cell\/}, {\bf 11}(12), 4241--4257.

\bibitem[Ghahramani(1998)Ghahramani]{ghahramani1998learning}
Ghahramani, Z. (1998).
\newblock Learning dynamic bayesian networks.
\newblock {\em Adaptive Processing of Sequences and Data Structures\/}, pages
  168--197.

\bibitem[Guo {\em et~al.}(2011)Guo, Levina, Michailidis, and Zhu]{guo2011joint}
Guo, J., Levina, E., Michailidis, G., and Zhu, J. (2011).
\newblock Joint estimation of multiple graphical models.
\newblock {\em Biometrika\/}, {\bf 98}(1), 1.

\bibitem[Husmeier(2003)Husmeier]{husmeier2003sensitivity}
Husmeier, D. (2003).
\newblock Sensitivity and specificity of inferring genetic regulatory
  interactions from microarray experiments with dynamic bayesian networks.
\newblock {\em Bioinformatics\/}, {\bf 19}(17), 2271--2282.

\bibitem[Inuzuka {\em et~al.}(2011)Inuzuka, Shaik, Onoyama, Gao, Tseng, Maser,
  Zhai, Wan, Gutierrez, Lau, {\em et~al.}]{Inuzuka11}
Inuzuka, H., Shaik, S., Onoyama, I., Gao, D., Tseng, A., Maser, R., Zhai, B.,
  Wan, L., Gutierrez, A., Lau, A., {\em et~al.} (2011).
\newblock Scffbw7 regulates cellular apoptosis by targeting mcl1 for
  ubiquitylation and destruction.
\newblock {\em Nature\/}, {\bf 471}(7336), 104--109.

\bibitem[Jensen(1996)Jensen]{jensen1996introduction}
Jensen, F. (1996).
\newblock {\em An introduction to Bayesian networks\/}, volume~74.
\newblock UCL press London.

\bibitem[Lauritzen(1996)Lauritzen]{lauritzen1996graphical}
Lauritzen, S. (1996).
\newblock {\em Graphical models\/}, volume~17.
\newblock Oxford University Press, USA.

\bibitem[McGlohon and Faloutsos(2007)McGlohon and Faloutsos]{mcglohon2007adage}
McGlohon, M. and Faloutsos, C. (2007).
\newblock {\em ADAGE: A software package for analyzing graph evolution\/}.
\newblock Carnegie Mellon University, School of Computer Science, Machine
  Learning Dept.

\bibitem[Meinshausen and B{\"u}hlmann(2006)Meinshausen and
  B{\"u}hlmann]{meinshausen2006high}
Meinshausen, N. and B{\"u}hlmann, P. (2006).
\newblock High-dimensional graphs and variable selection with the lasso.
\newblock {\em The Annals of Statistics\/}, {\bf 34}(3), 1436--1462.

\bibitem[Meinshausen and B{\"u}hlmann(2010)Meinshausen and
  B{\"u}hlmann]{meinshausen2010stability}
Meinshausen, N. and B{\"u}hlmann, P. (2010).
\newblock Stability selection.
\newblock {\em Journal of the Royal Statistical Society: Series B (Statistical
  Methodology)\/}, {\bf 72}(4), 417--473.

\bibitem[Murphy(2002)Murphy]{murphy2002dynamic}
Murphy, K. (2002).
\newblock {\em Dynamic bayesian networks: representation, inference and
  learning\/}.
\newblock Ph.D. thesis, University of California.

\bibitem[Natarajan {\em et~al.}(2001)Natarajan, Meyer, Jackson, Slade, Roberts,
  Hinnebusch, and Marton]{natarajan2001transcriptional}
Natarajan, K., Meyer, M., Jackson, B., Slade, D., Roberts, C., Hinnebusch, A.,
  and Marton, M. (2001).
\newblock Transcriptional profiling shows that gcn4p is a master regulator of
  gene expression during amino acid starvation in yeast.
\newblock {\em Molecular and Cellular Biology\/}, {\bf 21}(13), 4347.

\bibitem[Nau {\em et~al.}(2002)Nau, Richmond, Schlesinger, Jennings, Lander,
  and Young]{nau2002human}
Nau, G., Richmond, J., Schlesinger, A., Jennings, E., Lander, E., and Young, R.
  (2002).
\newblock Human macrophage activation programs induced by bacterial pathogens.
\newblock {\em Proceedings of the National Academy of Sciences\/}, {\bf 99}(3),
  1503.

\bibitem[Neapolitan(2004)Neapolitan]{neapolitan2004learning}
Neapolitan, R. (2004).
\newblock {\em Learning bayesian networks\/}.
\newblock Pearson Prentice Hall Upper Saddle River, NJ.

\bibitem[Perrin {\em et~al.}(2003)Perrin, Ralaivola, Mazurie, Bottani, Mallet,
  and d’Alche Buc]{perrin2003gene}
Perrin, B., Ralaivola, L., Mazurie, A., Bottani, S., Mallet, J., and d’Alche
  Buc, F. (2003).
\newblock Gene networks inference using dynamic bayesian networks.
\newblock {\em Bioinformatics\/}, {\bf 19}(suppl 2), ii138.

\bibitem[Preiss(2008)Preiss]{preiss2008data}
Preiss, B. (2008).
\newblock {\em Data structures and algorithms with object-oriented design
  patterns in C++\/}.
\newblock A1bazaar.

\bibitem[Rangel {\em et~al.}(2004)Rangel, Angus, Ghahramani, Lioumi, Sotheran,
  Gaiba, Wild, and Falciani]{rangel2004modeling}
Rangel, C., Angus, J., Ghahramani, Z., Lioumi, M., Sotheran, E., Gaiba, A.,
  Wild, D., and Falciani, F. (2004).
\newblock Modeling t-cell activation using gene expression profiling and
  state-space models.
\newblock {\em Bioinformatics\/}, {\bf 20}(9), 1361--1372.

\bibitem[Tibshirani(1996)Tibshirani]{tibshirani1996regression}
Tibshirani, R. (1996).
\newblock Regression shrinkage and selection via the lasso.
\newblock {\em Journal of the Royal Statistical Society. Series B
  (Methodological)\/}, pages 267--288.

\bibitem[Wang {\em et~al.}(2009)Wang, Sun, and Toh]{wang2009solving}
Wang, C., Sun, D., and Toh, K. (2009).
\newblock Solving log-determinant optimization problems by a newton-cg primal
  proximal point algorithm.
\newblock {\em preprint\/}.

\bibitem[Whittaker(1990)Whittaker]{whittaker1990graphical}
Whittaker, J. (1990).
\newblock {\em Graphical models in applied multivariate statistics\/},
  volume~16.
\newblock Wiley New York.

\bibitem[Wit and Abbruzzo(2012)Wit and Abbruzzo]{wit2012}
Wit, E. C. and Abbruzzo, A. (2012).
\newblock Factorial graphical lasso for dynamic networks.
\newblock Technical report.

\end{thebibliography}

\end{document}